\documentclass[12pt]{article} 
\usepackage{hyperref} 
\usepackage{geometry} % see geometry.pdf on how to lay out the page. There's lots.

\pdfoutput=1 
\title{Self Healing Soap Films} 
\author{Entry\#: 102387\\
Taylor Killian, Jordan Huey, Joshua Bryson \& Tadd Truscott \\
\\
Department of Mechanical Engineering \\ 
Brigham Young University, Provo, UT, 84602 USA} 

\date{}

\begin{document} 
\maketitle 
%% The abstract (in this file, and that submitted as text to arXiv) should include the exact phrase 
%% "fluid dynamics video" or "fluid dynamics videos" 
\begin{abstract} 
In 1904, while experimenting with high-speed photography, Lucien Bull recorded a pellet passing through a soap bubble. We investigate the dynamics that allow for a rigid body to pass through a hemispherical soap film without rupturing it. In this fluid dynamics video spheres were dropped from rest above a hemispherical soap film. At impact, the soap film stretches into a cavity around the sphere. As the sphere continues to descend, the film cavity pinches off and the film returns to its initial hemispherical shape. Upon closer observation of the film-sphere-air interface, the stability of the soap film appears to arise through a balance between the forces of the sphere inertia and the film tension. Therefore the relevant experimental parameter is the Weber number: $We=\frac{\rho2ghR}{\sigma}$, where $R$ is the sphere radius and $h$ is the height that the sphere is dropped from. We vary the sphere radius and velocity to provide a range of Weber numbers in order to investigate the dependence of film stability. Three subtly distinct regimes arise across the observed experimental range of Weber numbers. The first regime is demonstrated ($We<3200$) by a cavity shape that resembles a catenoid which pinches off both near the sphere and the hemispherical surface. The second regime is demonstrated ($3200<We<6100$) by a uniformly narrowing cavity resembling an inverted cone that pinches off near the sphere. The third regime initially appears similar to the second but the greater sphere inertia ($6100<We$) elongates the cavity shape and causes pinch-off and cavity collapse to be non-uniform in nature. \cite{Bryson2011}

%$We=\frac{\rho U^2R}{\sigma}=\frac{\rho2ghR}{\sigma}$, where U is the sphere velocity ($U = \sqrt{2gh}$), $R$ is the sphere radius, $g$ is the gravitational constant and $h$ is the height that the sphere is dropped from. We vary the sphere radius and velocity to provide a range of Weber numbers in order to investigate the dependence of film stability on Weber number. Three subtly distinct regimes arise across the observed experimental range of Weber numbers ($1500\leq We \leq13500$). The first regime is demonstrated ($We<3070$) by a cavity shape that resembles a catenoid which pinches off both near the sphere and the hemispherical surface. The second regime is demonstrated ($3070<We<6090$) by a uniformly narrowing cavity resembling an inverted cone that pinches off near the sphere. The third regime initially appears similar to the second but the greater sphere inertia ($6090<We$) elongates the cavity shape and causes pinch-off and cavity collapse to be non-uniform in nature.
\end{abstract}
% main text 
%\section{Introduction} 
%The {\em hyperref} package is used to make links to the videos. 
%%%  The format is:  \href{URL of video}{name that will appear in the text} 
%Two sample videos are 
%\href{http://ecommons.library.cornell.edu/bitstream/1813/8237/2/LIFTED_H2_EMS
%T_FUEL.mpg}{Video 
%1} and 
%\href{http://ecommons.library.cornell.edu/bitstream/1813/8237/4/LIFTED_H2_IEM
%_FUEL.mpg}{Video 
%2}. 
%It is recommended that the article include: 
%\begin{enumerate} 
%\item An explanation of what is shown in the video. 
%\item The relevant conditions, parameters, etc.. 
%\item References to any papers containing further information on the 
%videos. 
%\item In the Abstract (in the LaTeX file and in the text submitted 
%to arXiv), the exact phrase ``fluid dynamics video" or ``fluid 
%dynamics videos".  This is to facilitate subsequent searching. 
%\end{enumerate} 
% 
\bibliographystyle{abbrv}
\bibliography{bib}

\end{document}